\documentclass[preprint,journal]{vgtc}       % preprint (journal style)

\ifpdf%                                % if we use pdflatex
  \pdfoutput=1\relax                   % create PDFs from pdfLaTeX
  \usepackage{graphicx}                % allow us to embed graphics files
  \DeclareGraphicsExtensions{.pdf,.png,.jpg,.jpeg} % for pdflatex we expect .pdf, .png, or .jpg files
\else%                                 % else we use pure latex
  \usepackage{graphicx}                % allow us to embed graphics files
  \DeclareGraphicsExtensions{.eps}     % for pure latex we expect eps files
\fi%

\graphicspath{{figures/}{pictures/}{images/}} % where to search for the images

\usepackage{microtype}                 % use micro-typography (slightly more compact, better to read)
\PassOptionsToPackage{warn}{textcomp}  % to address font issues with \textrightarrow
\usepackage{textcomp}                  % use better special symbols
\usepackage{mathptmx}                  % use matching math font
\usepackage{times}                     % we use Times as the main font
\usepackage{cite}                      % needed to automatically sort the references
\usepackage{tabu}                      % only used for the table example
\usepackage{booktabs}                  % only used for the table example
\usepackage{amsmath}
\usepackage{amssymb}

\newcommand{\len}{l}
\newcommand{\err}{\epsilon}
\newcommand{\pp}{\tilde{x}_k} %Portal Point, string-tightened point on portal
\onlineid{1139}

\vgtccategory{Research}
\vgtcpapertype{algorithm/technique}

\title{Dynamic Portal Occlusion for Precomputed Interactive Sound Propagation}

\author{Nikunj Raghuvanshi}
\authorfooter{
\item
 Nikunj Raghuvanshi is with Microsoft Research. E-mail: nikunjr AT microsoft.com.
}

\shortauthortitle{Raghuvanshi: Dynamic portals for sound propagation}
\abstract{
An immersive audio-visual experience in games and virtual reality requires fast calculation of diffraction-based acoustic effects. To maintain plausibility, the effects must retain spatial smoothness on source and listener motion within geometrically complex scenes. Precomputed wave-based techniques can render such results at low runtime CPU cost, but remain limited to static scenes. Modeling the occlusion effect of dynamic portals such as doors present an unresolved challenge to maintain audio-visual consistency. We present a fast solution implementable as a drop-in extension to existing precomputed systems. Key is a novel portal-search method that leverages precomputed propagation delay and direction data to find portals intervening the diffracted shortest path connecting dynamic source and listener at runtime. The method scales linearly with number of portals in worst case, far cheaper than explicit global path search that scales with scene area. We discuss culling techniques to accelerate further. The search algorithm is combined with geometric-acoustic approximations to model the additional direct and indirect energy loss from intervening portals depending on their dynamic closure state. We demonstrate plausible audio-visual animations within our system integrated with Unreal Engine 4 (TM) and Wwise (TM).
} % end of abstract


\CCScatlist{ 
 \CCScat{I.3.7}{Computer Graphics}{Three-Dimensional Graphics and Realism}{Virtual reality};
 \CCScat{K.7.m}{Information Interfaces and Presentation}{Sound and Music Computing}{Modeling}
}

\vgtcinsertpkg

\begin{document}
%Precomputed Practical and in current usage.
%Simulating with open and close combinatorial explosion.
%why not precompute this as well - find portals beforehand? MULtiple portals .
\firstsection{Introduction}
\maketitle
Audio-visual consistency and immersion in games and virtual reality (VR) requires fast modeling of sound propagation. Typical scenes in these applications are geometrically and topologically complex, with millions of triangles and hundreds of spaces connected by portals and passages. Rendering plausible acoustic effects that vary smoothly and consistently in such scenes require expensive wave diffraction modeling, to capture sound redirection and occlusion around portals, among other wave effects. At the same time, a practical rendering must take a fraction of a single CPU core while updating acoustic effects (ideally) at visual frame rates to ensure audio-visual synchronization. 

Precomputed approaches tackle this challenge by assuming a static scene to shift the expensive wave simulation to an offline simulation and encoding stage, enabling their current use in major games \cite{Raghuvanshi_ASA_Triton:2017} and VR \cite{Godin_TritonWindowsMR:2018}. However, dynamic scene elements violate the static scene assumption. Portals such as doors and windows are the most commonplace dynamic element, currently typically modeled as fully open. This results in distracting audio-visual inconsistencies  when portals close visually yet don't cause progressive occlusion for sounds on the other side. Simple extensions like precomputing with portals open and closed suffers from combinatoric explosion of precomputation time and stored data. 

Dynamic portal occlusion is thus a major open problem that is of immediate practical importance for immersive audio-visual rendering in VR. In this paper, we make progress in this direction, showing dynamic portal effects within a pre-computed wave-based system within a practical CPU budget. Our method is formulated as a self-contained extension to existing systems; we demonstrate by extending \cite{Raghuvanshi:2018:Triton}. Given a dynamic source and listener location at runtime, we identify the intervening portals on the perceptually-salient initial sound as it undergoes diffracted shortest-path propagation from source to listener. Our main technical contribution is a fast search method for identifying the intervening portals, by leveraging geometric properties of the precomputed wave propagation delay and direction data. 

Our search method's CPU cost is worst-case linear in the number of portals, substantially faster than global path search such as with flooding on a grid. Culling tests are proposed that rule out most portals in practice, resulting in substantial further acceleration. We then employ geometric-acoustic assumptions to model the additional energy loss from intervening portals based on their current closure state for both initial and reflected (indirect) sound at the listener for each sound source. Overall, our technique is lightweight and generates plausible renderings, which we demonstrate with an integration in Unreal Engine 4 and AudioKinetic Wwise in a variety of complex game scenes.

\section{Related Work}
\textbf{Geometric acoustic systems} use high-frequency approximation to trace rays of sound. They have been extensively studied in the room acoustics community, see \cite{SaviojaGASurvey:2015} for a survey and current challenges. Real-time auralization in the community has focused on architectural walk-through applications where full use of computational resources of a single or few machines for audio rendering purposes is acceptable \cite{schroeder:2011}. General modeling of diffraction remains a tough open problem.

For interactive gaming and VR applications, geometric methods offer the promise of online simulation with dynamic scene geometry but add the additional challenge of fitting the extensive path exploration within fraction-of-core audio CPU budgets. Insufficient path tracing results can surprise the user with sudden changes on source or listener motion, an issue whose systematic study has recently started \cite{Cao_PathTracingInteractiveAcoustics:2016}. Commercial geometric-acoustic systems for games ignore diffraction entirely, requiring the user to choose detailed algorithmic parameters to ensure robust results \cite{SteamAudio}. The lack of portal diffraction modeling can cause unbounded errors in initial sound direction and loudness that is perceptually critical for spatially- and visually-consistent auditory localization \cite{litovsky:99}. 

\textbf{Interactive wave-based systems} \cite{raghuvanshi:2014, Raghuvanshi:2018:Triton, ravish:2013, Chaitanya_DirectionalSources:2020}, in contrast to geometric acoustics, employ first-principles solution of the fundamental acoustic wave equation and can thus naturally model diffraction directly on arbitrarily complex scenes. Assuming a static scene, they precompute the expensive wave simulation and then compress it using perceptual \cite{Raghuvanshi:2018:Triton} or physical \cite{ravish:2013} considerations. Runtime expense is thus reduced to decompression and lookup in an acoustic dataset for the scene. Research over the past decade has allowed such techniques to model moving sources and listener, but so far they are limited to static scenes, which we partly address in this paper by incorporating dynamic portals.

Past work has considered modular approaches similar in spirit to ours that separately account for the perturbation due to dynamic scene elements on global energy transport. Tsingos et. al. \cite{TsingosGascuel_FresnelZoneOcclusion:1997} introduced the idea of using Fresnel Zones to approximate smooth diffracted occlusion from dynamic occluders well-separated from the scene (such as fences or boulders outdoors) in a geometric-acoustic simulation otherwise lacking diffraction. The idea was recently refined in \cite{Runta_DiffractionKernels:2018}. Our interest is in the complementary problem of modeling portal occlusion. A more recent approach proposes real-time wave simulation on dynamic scene geometry \cite{Rosen_Planeverb:2020} but remains limited to only simulating a 2D slice of the scene rather than full 3D propagation of interest here.

The idea of combining the effect of multiple portals in series has been studied in  \cite{Stavrakis_ReverbGraphs:08}. The user is required to manually construct a room-portal decomposition of the scene that is used to construct a topopogical graph of the scene with rooms as vertices and portals as edges. Transfer properties within each room are pre-computed using geometric acoustics. At runtime, given any source and listener location, the detailed reverberant energy decay curve can be computed by using graph search to find connecting topological paths and modularly combining the pre-computed room and portal responses on each path. With our approach, such manual decomposition is unnecessary since a global wave solution with all portals open is already computed via simulation, but we borrow the general idea that reverberant energy transport through a series of portals along a path can be plausibly combined based on diffuse-field energy considerations.

\textbf{Path-finding methods} have a rich history in robot motion planning. More relevant to our work are methods commonly used in game AI. Our method is faster because it is precomputed, trading runtime computation for memory use. In fact, instead of a wave simulation, one could compute initial delay and direction information using Fast Marching Method or A* and our technique could then be combined with that data, as they correspond to an approximate first-wavefront wave simulation. %This could yield an efficient precomputed path-finding system, which is an avenue for future work. %The essence of what we're saying is: what kind of precomputed data do you store so you don't store too much and can find portals quickly at runtime? Our answer is: just delay and direction data - not complete path metadata. 

%Applying game AI methods to our problem: expensive and won't work - our source can be in 3D anywhere and listener is in 2D. game AI tackles both start and stop points in 2D on nav mesh. Potential application is the opposite: our method could be used to help game AI find the nearest door to get to listener leveraging precomputed audio data.

\begin{figure}
    \centering
    \includegraphics[width=\linewidth,trim={0 2.3in 3.5in 0},clip]{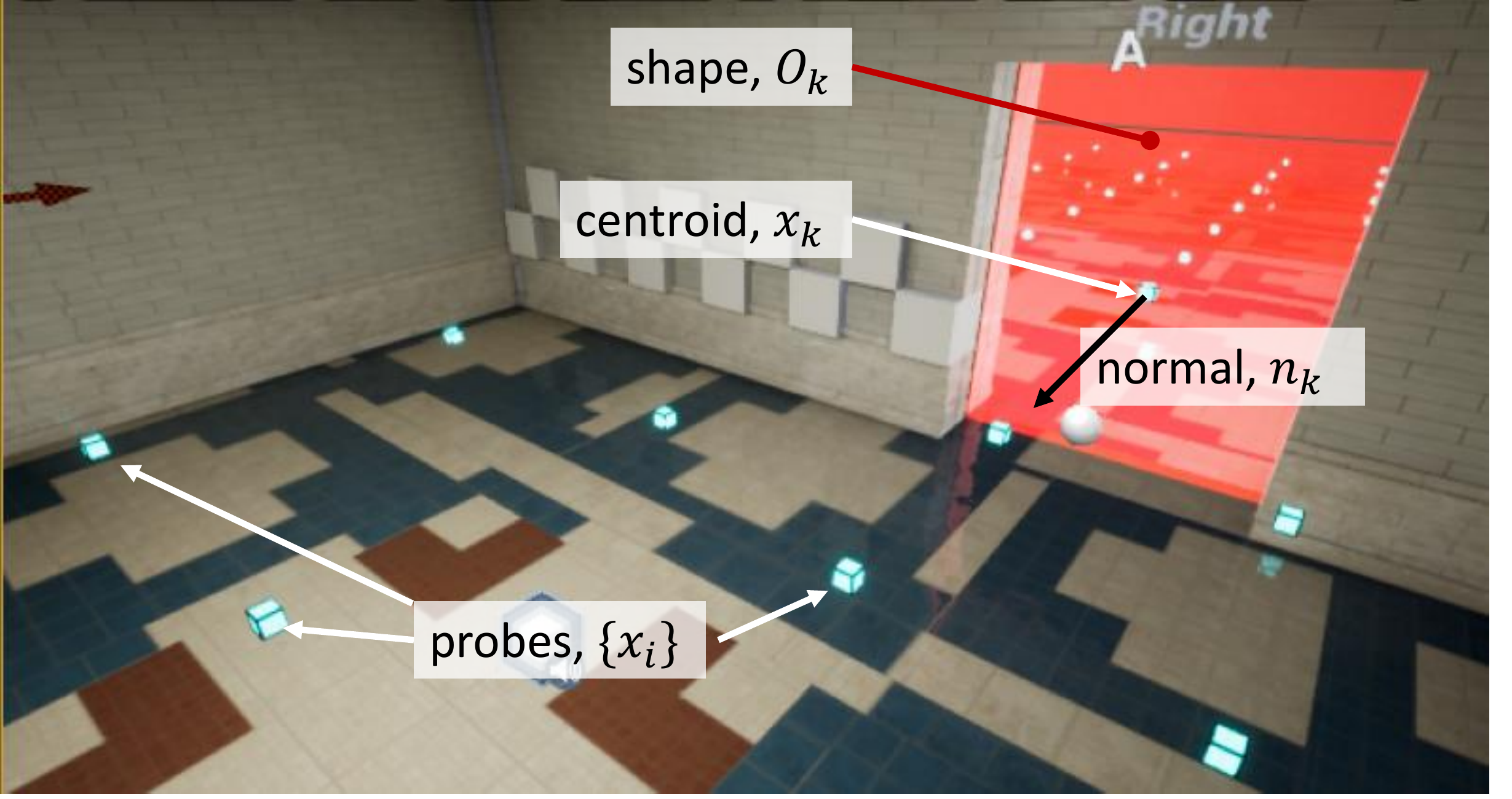}
    \caption{Portal markup and related geometric properties.}
    \label{fig:markup}
\end{figure}

\section{Background}
\label{sec:background}
%Source can fly in 3D. 6D problem. Probe-to-probe based chaining will not work.

We briefly review details from \cite{Raghuvanshi:2018:Triton} relevant to our work. The framework first computes a set of ``probe'' locations $\{x_i\}$ that sample the space of potential locations the listener might visit at runtime. The probe locations are unstructured, laid out above walkable regions in the scene using an adaptive sampling approach \cite{AllaChaitanya_AdaptiveSamplingTriton:2019} with variable spacing of 0.5-4m. Acoustic reciprocity is employed by treating each $x_i$ as a \emph{source} during pre-computation. Each probe computes a volumetric wave simulation on a cuboidal region centered at the probe, using a wave solver \cite{raghuvanshi_ARD:09}. The simulation grid is sub-sampled on a uniform 3D ``emitter grid'' of points $\{x_j'\}$ that represent potential runtime source locations, with a spacing of 1-1.5m. The directional impulse response at each simulation grid cell is passed through a streaming encoder that extracts perceptual acoustic parameters. The overall simulation process thus results in a set of perceptual acoustic parameters (discussed below) that depend on runtime source and listener location pair, $(x_j',x_i)$. The data is quantized and losslessly compressed. At runtime, given a source-listener location pair $(x',x)$, the parameters are decompressed, de-quantized, and spatially interpolated using nearby points $(x_j',x_i)$. This decoding is the primary CPU cost. The two main sources of error are quantization and spatial interpolation.

The encoded parameters relevant to the current paper are as follows. They collectively capture the time of arrival and distribution of energy around the listener at location $x$ due to a source at $x'$.

\begin{itemize}
    \item \textbf{Initial delay, $\tau_0$.} Assuming the impulsive source goes off at $t=0$, the wavefront propagating from source to listener $x'\rightsquigarrow x$ along shortest path through the scene, potentially diffracting around intervening portals, first arrives at time $t=\tau_0$. Values are quantized at 2ms. The length of the shortest path can be computed as $c\tau_0$, where $c=340$m/s is the speed of sound.
    
    \item \textbf{Initial loudness, $L$.} The initial sound is defined as energy arriving immediately after $\tau_0$ within a 10ms period, that is perceptually fused into a single, perceptually-dominant acoustic event due to the Precedence Effect \cite{litovsky:99}. The corresponding loudness is $L$ in dB.
    
    \item \textbf{Initial direction, $s_0$.} The propagation direction of the initial sound wavefront as it crosses the listener. For example, if the sound arrives at the listener through a visible portal, this direction will point from the portal towards the listener.
    
    \item \textbf{Directional reflections loudnesses, $R_J$.} Indirect reflected energy is accumulated in a 80ms time window. Its spherical energy distribution around the listener is represented using a linear combination of Cosine-squared basis functions centered on the six axial directions $S_J=\{+Z,+X,+Y,-X,-Y,-Z\}$ in world space. The weight for each basis function, $R_J$ corresponds to indirect energy arriving around axial direction $S_J$ converted to logarithmic (dB) scale.   
\end{itemize}

\section{Overview}
Our overall approach is to employ the acoustic parameters to identify portals that participate as initial sound propagates from $x'$ to $x$ (\autoref{sec:PortalSearch}), and then appropriately modify the energetic parameters $\{L,R_J\}$ to model the additional occlusion introduced by the portals (\autoref{sec:PortalOcclusion}). From there the rendering process remains identical to \cite{Raghuvanshi:2018:Triton}. In this section we first briefly summarize the overall architecture.

\subsection{Precomputation}
\label{sec:baking}
We use $k\in[1,N]$ to index over portals. Let $O_k$ denote the convex polygon representing the k-th portal opening, $n_k$ its normal, $x_k$ its centroid, and $r_k$ the radius of a bounding circle with center $x_k$. Our method requires that portal openings be explicitly marked within the 3D scene as convex 2D polygons. In typical game editor workflows this need not involve manually marking each portal. It is typical for portals such as doors and windows to be implemented as instances of a few archetype classes that encapsulate the visual, auditory, animation, and interaction properties of the object class (called ``Blueprint Actors'' in Unreal Engine and ``Prefabs'' in Unity Engine).  This allows editing on each portal type to be automatically propagated to all actual instances in the virtual world. An example markup is illustrated in \autoref{fig:markup}.

Once the additional portal markup has been performed the geometric portal properties described above are collected. For reasons to be described shortly, we perform additional simulations from the center of each portal, thus augmenting the set of simulated probes to $\{x_i\}\cup\{x_k\}$. From there, precomputation proceeds the same as \cite{Raghuvanshi:2018:Triton}, as summarized in \autoref{sec:background}. All portals are kept open during precomputation to ensure energy is able to propagate throughout the scene.

\subsection{Runtime}
Precomputation passes the portal information collected above to the runtime component, in addition to the encoded perceptual parameter dataset that is decoded at runtime as described in \cite{Raghuvanshi:2018:Triton} and summarized in \autoref{sec:background}. We will use the notation $P(x_{source},\,x_{listener})$ to denote lookup of parameter $P$ with first argument as source location and second argument listener location. Runtime computation proceeds in two steps discussed in sections \ref{sec:PortalSearch} and \ref{sec:PortalOcclusion}. First, given the source and listener locations $(x',x)$, search for all portals that intersect the initial sound propagating from source to listener. Second, we use the dynamic state for the intervening portals to modify the energetic acoustic parameters to render additional occlusion.

\section{Portal Search}
\label{sec:PortalSearch}
We present our central idea for quickly finding intervening portals, before discussing necessary refinements. Assume for argument's sake that portal size can be ignored ($r_k\to 0$) and consider the $k$-th portal with centroid at $x_k$. Shortest paths obey optimal sub-structure: $x_k$ lies along the shortest path $x'\rightsquigarrow x$, if and only if $x'\rightsquigarrow x_k$ and $x_k\rightsquigarrow x$ are also shortest paths. Same applies to their path lengths. Observe that the precomputed initial sound delay, $\tau_0$, is proportional to the length of the shortest path via $c\tau_0$. Thus, finding all portals on the shortest path connecting $x'$ to $x$ amounts to a linear search to find all $k\in[1,N]$ such that,
\begin{equation}
    \tau_0(x',x) = \tau_0(x',x_k) + \tau_0(x, x_k).
    \label{eq:PortalFindTheory}
\end{equation}
Note that this test does not require any explicit construction of the sub-paths, for instance, by using an expensive A* march from $x'$ to $x$ on a grid, which in the worst-case scales with the area of the scene. Instead we leverage the extensive geometric exploration of scene topology already performed by wave simulations during precomputation. There can be multiple portals that satisfy this equation if they all lie along the shortest path.

In practice, we must account for two sources of deviation. First, portals have finite size so that the shortest path will pass through somewhere within the portal $O_k$ rather than its centroid $x_k$, and second, the shortest path delay estimates in $\tau_0$ contain errors from quantization and spatial interpolation. The rest of this section discusses refinements necessary for making this observation practical. 

\begin{figure}
    \centering
    \includegraphics[width=\linewidth,trim={0 2.3in 5.6in 0},clip]{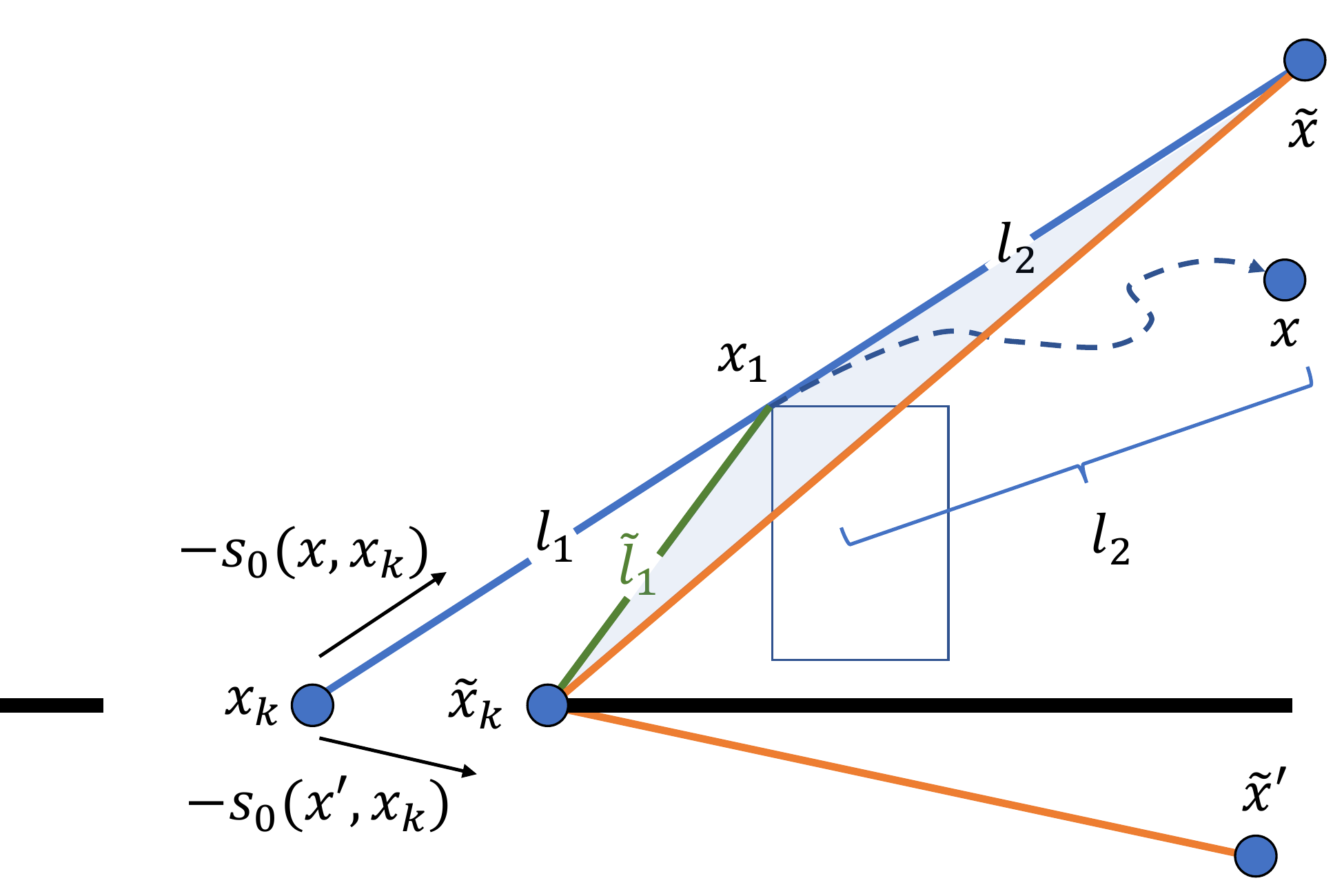}
    \caption{\textbf{Adjusting delay with portal shape.} Precomputed data is available only for portal centroid at $x_k$. We compute apparent source and listener locations $\{\tilde{x}',\tilde{x}\}$ and find the point $\tilde{x}_k$ where the ``string tightened'' path passes through the portal. Triangle inequality on shaded region shows that the corrected delay estimated this way provides a lower bound for actual propagation delay.}
    \label{fig:StringTighten}
\end{figure}

\subsection{Accounting for portal geometry}
\label{sec:StringTighten}
While we know the shortest path distance from source to listener $c\tau_0(x',x)$, we don't know which point on the portal $\tilde{x}_k \in O_k$ the path went through. For portals of finite size, the relation in \autoref{eq:PortalFindTheory} is accurate only with the substitution $x_k\leftarrow \pp$. We employ a ``string tightening'' procedure to estimate $\tilde{x}_k$ by exploiting the delay and direction data precomputed for the ``portal probe'' placed at the centroid, $x_k$. 

The method is illustrated in \autoref{fig:StringTighten}. Precomputed data for the probe is decoded to obtain the arrival delay $\tau_0(x,x_k)$ and propagation direction $s_0(x,x_k)$ of the initial sound at $x_k$, if a source had been placed at $x$. This is used to compute the apparent listener location,
\begin{equation}
    \tilde{x} = x_k - c\tau_0(x,x_k) s_0(x,x_k).
\end{equation}
The apparent source location is similarly computed,
\begin{equation}
    \tilde{x}' = x_k - c\tau_0(x',x_k) s_0(x',x_k) .
\end{equation}
We then find the point $\tilde{x}_k \in O_k$ that minimizes $|\tilde{x} - \tilde{x}_k| + |\tilde{x}' - \tilde{x}_k|$ to find the shortest path connecting $\tilde{x}'$ and $\tilde{x}$ via the portal, shown with orange lines in the figure. As we show next, it suffices to modify \autoref{eq:PortalFindTheory} to:
\begin{equation}
    |\tilde{x}'-\tilde{x}_k| + |\tilde{x}-\tilde{x}_k| \leq c\tau_0(x',x) 
    \label{eq:PortalFindFinite}
\end{equation}

To derive this upper bound, first consider the path from listener at $x$ up to the portal. %Note that all points are on the same plane except x.
Per \autoref{fig:StringTighten} the shortest path from $x_k$ to $x$ will consist of piece-wise linear segments as it wraps around scene geometry. Let the first segment's length be $l_1$, with end-point $x_1=x_k-l_1 s_0$. The remaining shortest path can be arbitrarily complex, shown with dashed line, with length $l_2 \geq 0$. By the definition of $\tilde{x}$, $x_k x_1 \tilde{x}$ is a line segment with length $l_1+l_2=c\tau_0(x,x_k)$. With portal centered at $x_k$, imagine growing its size smoothly from zero: the shortest path from $x$ to $x'$ will be topologically similar to the path going through $x_k$, differing only in going through the shorter segment $x_1 \tilde{x}_k$ with length $\tilde{l}_1$ with total shortest path length of $\tilde{l}_1 + l_2$ from listener up to the portal. This argument holds as long as the portal isn't large enough to cause a discontinuous jump in the shortest path from $x'$ to $x$. In practice, it works robustly for portals a few meters across such as doors and windows. 

Consider the shaded triangle $\tilde{x} x_1 \tilde{x}_k$ in \autoref{fig:StringTighten}. The triangle inequality shows that $|\tilde{x}-\tilde{x}_k|\leq\tilde{l}_1+l_2$. Applying the same argument from source to portal, $|\tilde{x}'-\tilde{x}_k|\leq\tilde{l}'_1+l_2'$. Summing these two inequalities, the right hand side becomes the shortest path length, yielding \autoref{eq:PortalFindFinite}. We emphasize that the path details such as $l_1$ and $l_2$ are used for the argument but are not available to the algorithm.

%How does the error vary? Goes to 0 when l_2 is 0. Its fine that it is a lower bound, but is there a bound on absolute error?

\begin{figure}
    \centering
    \includegraphics[width=\linewidth,trim={1.7in 0.5in 0.7in 0.5in},clip]{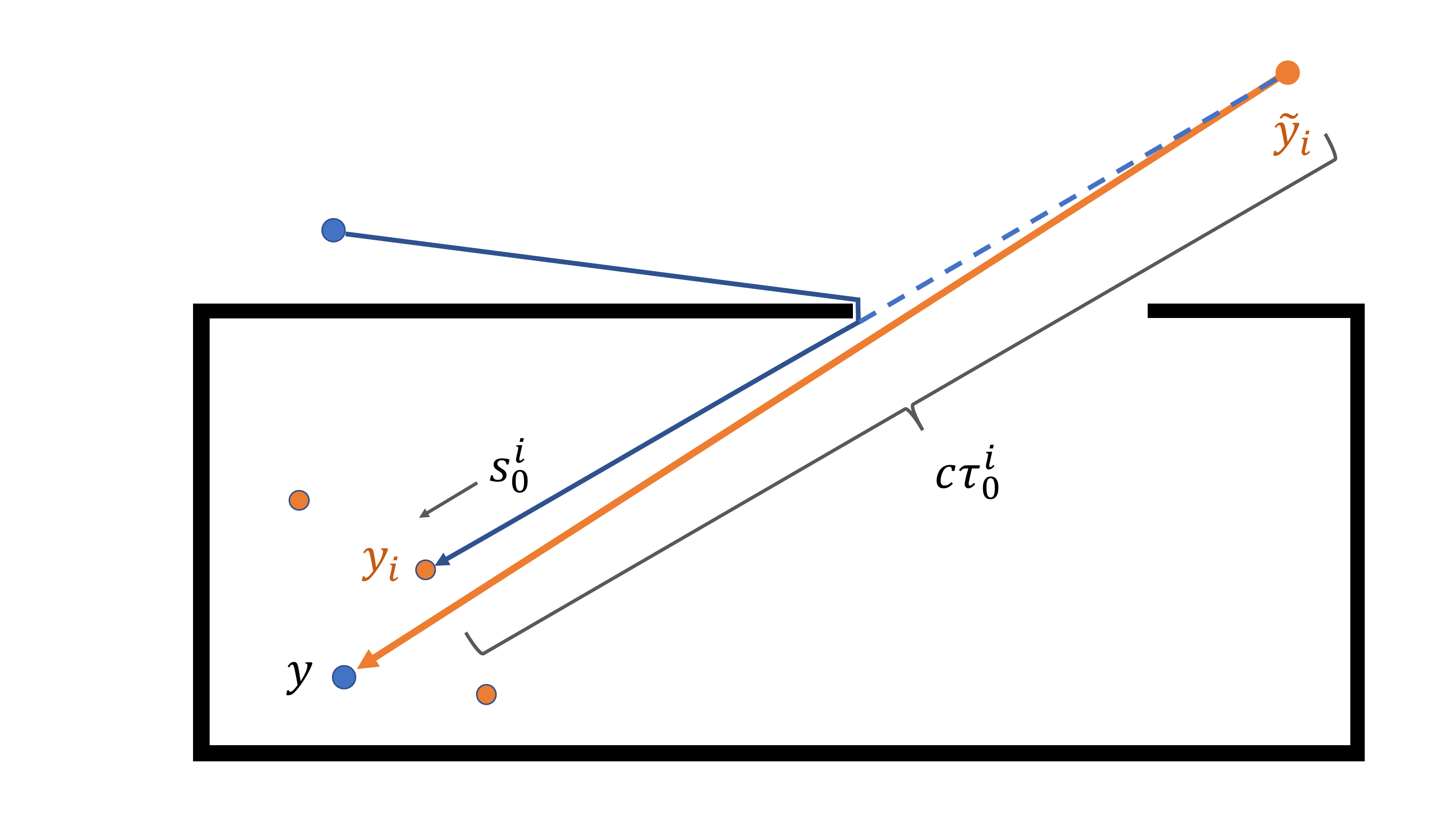}
    \caption{\textbf{Improved spatial interpolation.} Initial wavefront arrival time $\tau_0^i$ and direction $s_0^i$ at a sample point $y_i$ is used to compute an apparent location $\tilde{y}_i$ from which delay and direction at the evaluation point $y$ are evaluated. The procedure is repeated and linearly blended across samples (shown in orange). Our approach allows improved interpolation compared to direct linear interpolation of sample values, also affording some limited extrapolation like in the case shown.}
    \label{fig:Interpolation}
\end{figure}

\subsection{Reducing interpolation errors}
As noted in \autoref{sec:background}, the delay and direction of initial sound $\{\tau_0,s_0\}$ contain errors from spatial interpolation that must be taken into account for a robust implementation. We fold such errors into a tolerance term, $\err$, yielding the final criterion we employ for portal search: 
\begin{equation}
    |\tilde{x}'-\tilde{x}_k| + |\tilde{x}-\tilde{x}_k| \leq c\tau_0(x',x) + c\err
    \label{eq:PortalFindFinal}
\end{equation}

We found that the interpolation algorithm in \cite{Raghuvanshi:2018:Triton} required rather large values with $\err\approx25$ms due to the few-meter sample spacing used in practice. We take a couple measures to minimize interpolation error, keeping $\err$ small. Firstly, we use acoustic reciprocity appropriately in \autoref{eq:PortalFindFinal} to treat the portal centroid $x_k$ as the listener location in all evaluations. Since there is a probe exactly at $x_k$ during precomputation, probe interpolation error is avoided. But interpolation errors on the emitter grid $\{x_j'\}$ (discussed in \autoref{sec:background}) remain. Secondly, we propose an improved algorithm below which reduces interpolation errors in estimating delay and direction, allowing us to use the tighter tolerance of $\err=10$ms, corresponding to path length difference of 3.4m, with robust results in all our test cases.

As discussed in \autoref{sec:background}, spatial interpolation is performed for both the listener and the source using unstructured probe samples $\{x_i\}$ and uniform-grid emitter samples $\{x'_j\}$ respectively. For either case, assume we are given an arbitrary set of sample points $\{y_i\}$ with interpolation weights $\{w_i\}$ with the corresponding parameters $(\tau_0^i, s_0^i)$. The approach in \cite{Raghuvanshi:2018:Triton} linearly interpolates as $(\tau_0, s_0)=\sum_i w_i \times (\tau_0^i, s_0^i)$. We propose a non-linear approach motivated by similar geometric considerations as \autoref{sec:StringTighten}:
\begin{equation}
    \begin{aligned}
        \tilde{y}_i &= y_i - c \tau_0^i s_0^i \\
        \tau_0 &= \sum_i w_i |y - \tilde{y}_i| / c \\
        s_0 &= \sum_i w_i (y - y_i) / |y - \tilde{y}_i|
    \end{aligned}
    \label{eq:ImprovedInterp}
\end{equation}
As illustrated in \autoref{fig:Interpolation}, for each sample at $y_i$, an apparent location $\tilde{y}_i$ is computed that would result in the observed arrival delay and direction at $y_i$ assuming free space propagation. This apparent location is used to evaluate the delay and direction contribution at the evaluation point, $y$, for this sample, linearly blended across all closeby samples. The results are found to be much more accurate in practice allowing a much smaller error tolerance $\err$ as mentioned earlier. The improvement is especially important in cases where $y$ is not entirely surrounded by samples like the case shown in \autoref{fig:Interpolation}. Our approach is able to afford a degree of extrapolation, in this case correctly producing an increased delay estimate compared to any of the samples. This case is a common occurrence, when the listener or source occupies the space between a sample and a wall.

%Explain why this interpolation is so great. Use the Doors example.

\begin{figure}
    \centering
    \includegraphics[width=\linewidth,trim={0 5.4in 6in 0in},clip]{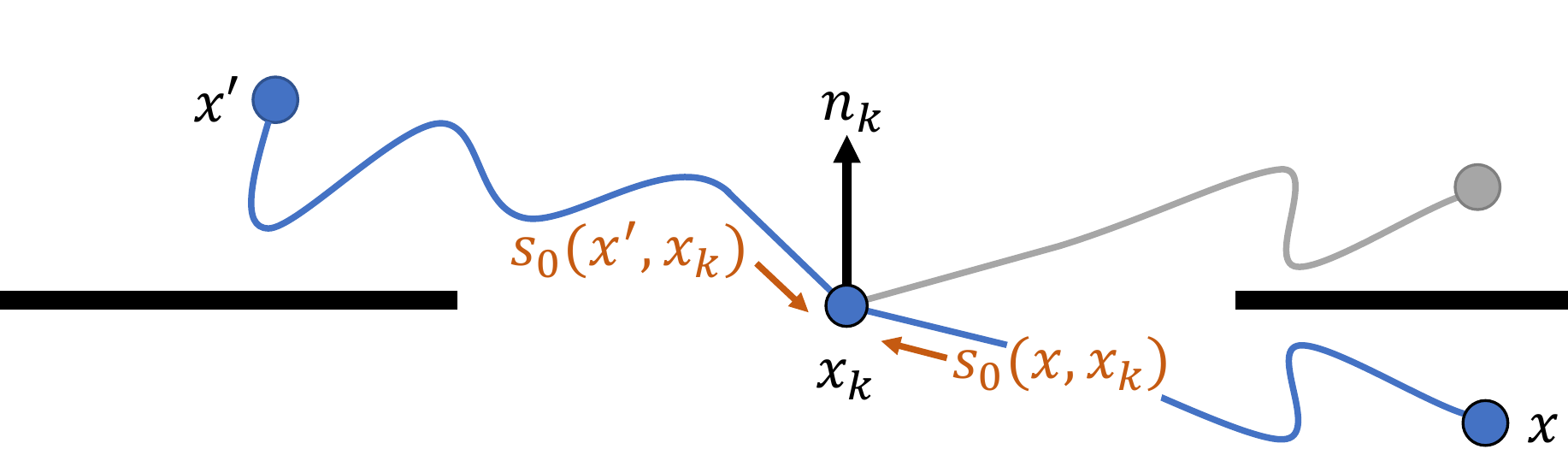}
    \caption{\textbf{Ensuring path pierces portal.} When the shortest path $x'\rightsquigarrow x$ passes close to a portal, the delay test in \autoref{eq:PortalFindFinal} can be erroneously satisfied to within the test's precision (gray). Treating portal centroid $x_k$ as a virtual listener, we check if the arriving wavefront at $x_k$ from $x$ and $x'$ are from opposing directions w.r.t the portal normal $n_k$, to allow only paths that actually pierce the portal (blue). }
    \label{fig:NormalTest}
\end{figure}

\subsection{Ensuring path pierces portal}
It is possible for the shortest path $x'\rightsquigarrow x$ to pass within the error tolerances inherent in \autoref{eq:PortalFindFinal} despite not piercing the portal. This case occurs frequently in practice when the source and listener are in a room and either of them walks close to a closed door. The result is immersion-breaking erroneous occlusion. We perform an additional test, illustrated in \autoref{fig:NormalTest}:
\begin{equation}
    s_0(x',x_k) \cdot n_k \times s_0(x,x_k) \cdot n_k < 0.
    \label{eq:PortalNormalCull}
\end{equation}
Treating portal centroid $x_k$ as listener we verify that the wavefront arriving at $x_k$ from $x$ and $x'$ are from opposite sides of the portal's plane, using its normal $n_k$. Note that negating $n_k$ preserves the result. As we show in our results, this test allows portal occlusion to respond robustly to motion of the source or listener as they approach and cross a portal at a much finer spatial resolution than the few meters used for acoustic sampling.

\subsection{Culling}
While evaluating the terms in \autoref{eq:PortalFindFinal} is much cheaper than global path search, each evaluation still requires decompression and interpolation of parameter fields  \cite{Raghuvanshi:2018:Triton}. This can become costly in practice as large scenes can contain hundreds of portals over kilometer-squared areas and such evaluations must be repeated per sound source. We propose some culling tests to reduce this cost.

Define the ellipsoidal volume $y\in \mathcal{E}(x',x)$ if $|y-x'| + |y-x| < \len_{max}$, with focii as $x$ and $x'$ and maximum path length $\len_{max}=c(\tau_0 + \err)$ on its surface. The interior of this ellipsoid contains all possible paths with length bounded as: $\len(x'\rightsquigarrow x) < \len_{max}$, a necessary condition for \autoref{eq:PortalFindFinal} to hold. Any portal that doesn't intersect this ellipsoid is too far away to participate. 

To further accelerate, we compute the ellipsoid's axis-aligned bounding box $B(x',x) \supset \mathcal{E}(x',x)$.  Using $B+r$ to denote $B$ enlarged by a scalar $r$ in all six axial directions, we accept a portal only if its centroid $x_k$ satisfies:
\begin{subequations}
\begin{align}
    x_k &\in B(x',x) + r_k \label{eq:PortalBboxCull}\\
    |x'-x_k| + |x_k-x| &< \len_{max} + r_k \label{eq:PortalEllipsoidCull}
\end{align}
\end{subequations}
The additional radius term $r_k$ conservatively accounts for portal size via a bounding sphere. The first bounding box test rejects most far-away portals and the second (more expensive) test checks for detailed ellipsoid intersection.

%Recall that precomputation adds a probe at each location $x_k$. Also notice that the last term has the arguments reversed so that probe centroid $x_k$ is treated as a virtual listener and the actual listener location $x$ as a source. This is arranged so required data for practical implementation of this equation is available as will clarify shortly.

\subsection{Algorithm summary}
For each source located at $x'$ with listener at $x$, the portal search algorithm iterates over all portal indices $k\in [1,N]$. For each, we first cull using \autoref{eq:PortalBboxCull} and then \autoref{eq:PortalEllipsoidCull}. If the tests pass, additional acoustic lookups are performed for $(x,x_k)$ and $(x',x_k)$. Then \autoref{eq:PortalNormalCull}, and \autoref{eq:PortalFindFinal} are applied in sequence and if satisfied, the portal is declared to lie on the initial sound path from source to listener and its occlusion processing is applied as described in the next section. Multiple portals can pass these tests in which case all their occlusion effects are accumulated. The output is a set of indices $K$ of portals lying on the shortest path, and also the index of the last portal along the path from source to listener,
\begin{equation}
    \kappa=argmin_{k\in K} |\tilde{x}-\tilde{x}_k|.
\end{equation}

\section{Portal Occlusion}
\label{sec:PortalOcclusion}
The essential approximation in our method is only modeling first-order perturbation on energy transfer across the portal as it closes dynamically, ignoring the additional loss on higher-order paths that cross the portal multiple times. For instance, consider two rooms with a shared door. The reverberation time in such cases can depend on the sizes of both rooms, which we do model since portals are left open during precomputation. However, as an acoustically-opaque door is shut, the reverberation times for the two rooms progressively decouple which is something we can't model. Similar considerations apply to the net indirect energy level at the listener. Limiting to small portals a few meters across is key to ensuring results from such an approach are plausible, capturing the primary audio-visual synchronization cues demonstrated in our results.

\subsection{Initial loudness}
Each portal dynamically maintains an open-area fraction, $\alpha_k \in [0,1]$ synchronized to visual animation. Since the precomputation already includes diffraction effects from portal geometry for $\alpha_k=1$, we only need to model the relative perturbation on initial loudness $L(x',x)$. We approximate this by assuming the energy transferred through a portal is proportional to its open area, invoking a geometric approximation. For the initial loudness this yields,
\begin{equation}
    L\leftarrow L + 10 \log_{10} \alpha,\text{ where }\alpha \equiv \prod_{k\in K}\alpha_k,
    \label{eq:DirectOcclusion}
\end{equation}
where we compute the initial sound's energy as $10^{L/10}$, multiply the accumulated fractional energy loss from all portals in $\alpha$ and transform back to perceptual dB domain. More sophisticated (and expensive) aperture loss diffraction models are certainly possible within our method, such as using a numerically-evaluated Kirchoff Approximation integral \cite{Tsingos_InstantSoundScatter:2007} over the detailed dynamic aperture geometry as it closes. Here we opted for simplicity and speed while retaining plausibility.

\subsection{Reflections loudness}
Accounting for portal occlusion on reflected sound is more involved because multi-bounce energy at the listener can also arrive via  other static openings in the world. Recall from \autoref{sec:background} that the spherical distribution of reflected loudness (in dB) at the listener is captured by parameters, $R_J(x',x)$ around six axial directions indexed by $J$. Define the corresponding directional energy as $E_*\equiv 10^{(R_*/10)}$, and total energy $E(x',x)=\sum_J E_J$. 

We first estimate $E_\kappa$, which is the portion of total energy $E$ that passes through the chain of portals $k\in K$ to finally exit from the last portal $\kappa$ to arrive at the listener. The total energy received on portal $\kappa$ from the source is approximately $A(O_\kappa)E(x',x_\kappa)$ where $A(O_\kappa)$ is the portal's area. This approximates by assuming a diffuse reverberant field with uniform energy distribution across the portal. To model the transport from the portal up to the listener, we imagine putting a source with matching radiated energy on the portal, and observe the reflected energy received at the listener. Point sources injected during simulation are normalized to have unit amplitude at 1m distance, corresponding to total radiated energy of $\int_{S^2} p^2 ds = 4\pi$. Invoking reciprocity so that $E(x_\kappa,x) = E(x,x_\kappa)$ to reduce interpolation error, we have the estimate, 
\begin{equation}
    E_\kappa(x',x) \approx \frac{A(O_\kappa)}{4\pi}E(x',x_\kappa) E(x,x_\kappa).
    \label{eq:EnergyThroughPortal}
\end{equation}
The true value of $E_\kappa$ must obey reciprocity since swapping $(x',x)$ reverses all propagation paths $x'\rightsquigarrow x$ which clearly represent the same net energy transfer across the portal. Note that our approximation preserves this important property, as it is manifestly invariant to interchanging $x$ and $x'$. 

Next we estimate how $E_\kappa$ gets directionally distributed around the listener after radiating from portal $\kappa$. We perform an additional acoustic lookup with portal as virtual source to compute $E_J(x_\kappa,x)$, and compute,
\begin{equation}
    E_J^\kappa(x',x) = \frac{E_\kappa(x',x)}{E(x_\kappa,x)}E_J(x_\kappa,x). 
\end{equation}
The first factor normalizes so that total energy $\sum_J E_J^\kappa = E_\kappa$ while preserving the relative directional distribution of reflected arrivals at the listener due to energy arriving from the portal, as captured in $E_J(x_\kappa,x)$. Thus, energy arriving in world direction $X_J$ at the listener can be split into that arriving through the portal of interest, $\kappa$: $E_J^\kappa(x',x)$, while the residual energy: $E_J(x',x) - E_J^\kappa(x',x)$, must arrive through other portals or openings in the environment. 

Assuming that the actual transmitted energy at runtime is in proportion to open area, same as in \autoref{eq:DirectOcclusion}, we get the net directional energy at listener due to portals on the shortest path: $\alpha E_J^\kappa(x',x)$. Summing with residual energy, we obtain the total directional energy at the listener: $E_J(x',x) - (1-\alpha) E_J^\kappa(x',x)$.
Finally, we update the reflection loudness parameters as,
\begin{equation}
    R_J \leftarrow 10\log_{10} \max\{\beta E_J,\,E_J - (1-\alpha) \,E_J^\kappa\}.
    \label{eq:ReflectedOcclusion}
\end{equation}
The clamp above $\beta E_J$ with $\beta = .01$ ensures we only remove up to $1-\beta=0.99$ fraction of the reflected energy. This ensures reverberation isn't entirely removed due to approximation inherent in \autoref{eq:EnergyThroughPortal} which can sound implausible.
%Have we seen it trigger frequently?
 
\section{Results and discussion}
\begin{figure}
    \centering
    \includegraphics[width=\linewidth,trim={0 3.8in 6in .5in},clip]{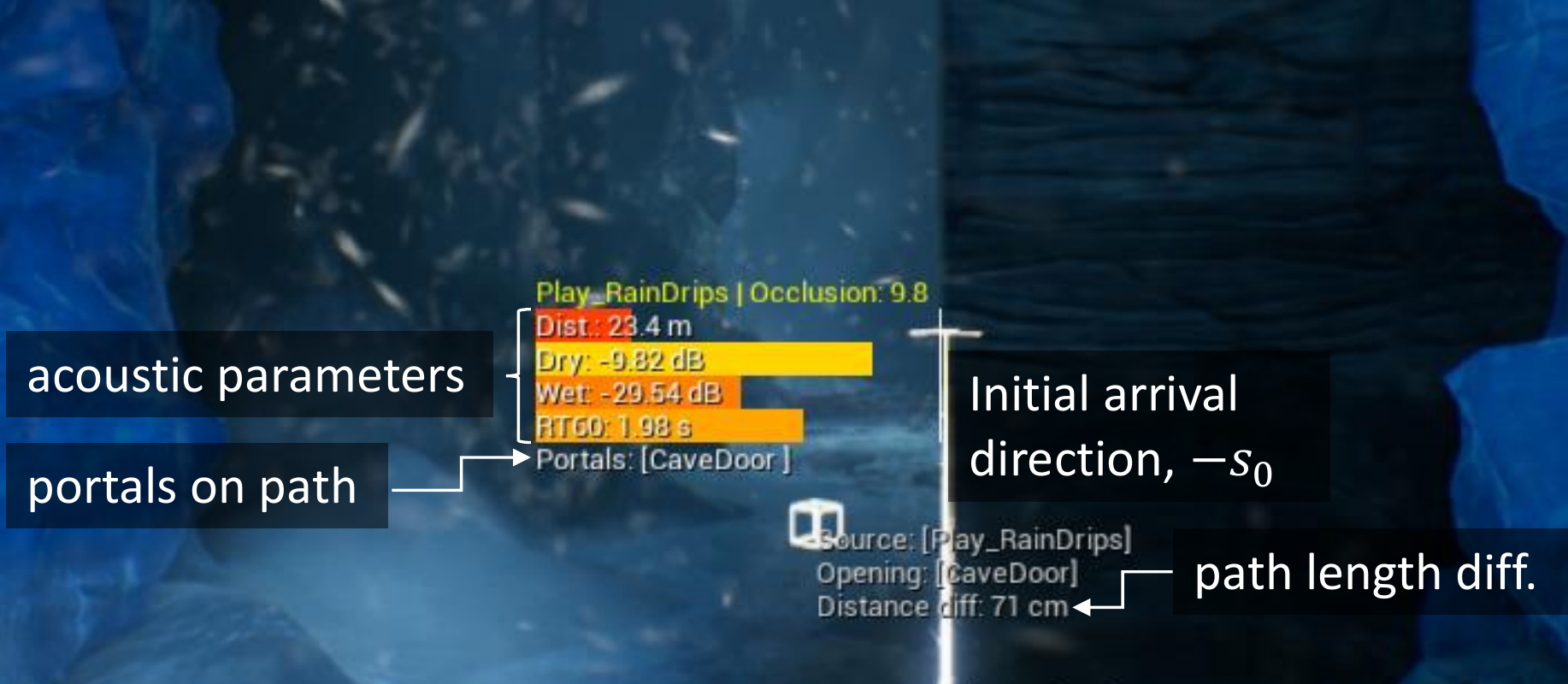}
    \caption{\textbf{Description of runtime values visualized in video.} Acoustic parameters are as follows: ``dist'' is the path length $c\tau_0$, ``dry'' is initial loudness $L$, ``wet'' is reflections loudness $10\log_{10}E$. Apparent source direction, $-s_0$, is shown with a white arrow emanating from the listener. Shortest path intersection with portal, $\pp$ is shown with a white box. ``Distance diff'' is the path length difference $|\tilde{x}'-\tilde{x}_k| + |\tilde{x}-\tilde{x}_k| - c\tau_0(x',x)$ which must be smaller than $c\epsilon$ to pass the portal inclusion test \autoref{eq:PortalFindFinal}.}
    \label{fig:debugvals}
\end{figure}

Our system is integrated with Unreal Engine 4 and AudioKinetic's Wwise audio engine.
In all cases we clamped the net energy loss factor $\alpha$ to a minimum of $.001$, corresponding to a maximum occlusion of $-30$dB. This was found to improve results to be in accord with everyday experience, presumably because forming a perfect acoustical seal is difficult and common doors have a degree of leakage. Please consult the accompanying video to hear the results discussed below with headphones. Runtime values shown in the video are explained in \autoref{fig:debugvals}. %We emphasize that our plausibility claims below are limited to the fact that as doors close, with our method the sound's loudness decreases. Whereas without such treatment, no such variation would occur.

In the \textsc{Courtyard} scene we have three portals with animated doors that are cyclically closing and opening. 
In the accompanying video, we show that as the listener moves around a room with source placed outside across a portal, the system consistently locates the single intervening portal. This shows that our portal search method in \autoref{eq:PortalFindFinal} works robustly in conjunction with improved interpolation in \autoref{eq:ImprovedInterp}. Without the latter, we observed that at specific listener locations the portal wouldn't be selected due to large interpolation errors in propagation delay violating \autoref{eq:PortalFindFinal}. This would then unexpectedly un-occlude audio despite a closed door. Such occlusion ``pops'' on source or listener motion are clearly intolerable for a practical system. Next we show that due to the path-piercing test in \autoref{eq:PortalNormalCull} we are able to model highly detailed occlusion changes as the listener crosses a closed door. 

We note that the initial sound's occlusion per energy considerations in \autoref{eq:DirectOcclusion} renders a plausible effect matching everyday experience: the loudness reduction is small initially, but reduces abruptly as the portal is about to close shut. We also demonstrate cases where there are multiple portals along the initial path, rendering their combined occlusion. 

The \textsc{Cave} scene demonstrates a typical use case in a geometrically complex scene. While the portal shape is irregular, note that any bounding convex polygon will satisfy \autoref{eq:PortalFindFinal}. We show a square shape for $O_k$ suffices which reduces the only manual markup work involved. We show that rain drip sounds inside the cave occlude naturally as the door closes with listener standing outside. We also illustrate with two sources how our technique enables a common sound design task of rendering the effect of a portal to mutually isolate an outdoor and indoor soundscape. Such effects are done by hand today. From everyday experience we have strong expectations of such acoustic effects, which our system is able to render automatically, efficiently, and plausibly.

\section{Conclusion and discussion}
We presented the first technique to model dynamic portals for interactive wave-based sound propagation, filling an important gap in the audio-visual plausibility of current precomputed systems. We proposed a novel linear-time search algorithm for portals along the shortest diffracted path from source to listener that re-uses precomputed arrival delay and direction data, avoiding costly explicit path search. We discussed acceleration techniques and improvements required in practice for robust behavior. Although our examples were limited to doors, our technique generalizes to any dynamic aperture whose location is known beforehand. For instance, if there is a small portion of wall in an otherwise static scene that can collapse during gameplay, it can effectively be treated as a dynamic portal.

A key limitation of our approach is that it cannot model cases with multiple portals in parallel connecting the source and listener with nearly same initial path length. In such cases, when closing a portal on the shortest path, we over-estimate the initial sound occlusion, and when closing portals on the nearly-shortest path, no occlusion will be applied. 

Along with addressing such limitations in the future, we also wish to explore potential applications of the presented ideas in path-planning applications for game AI since our approach avoids explicit path search which is a major cost of current approaches.

\bibliographystyle{abbrv-doi}

\bibliography{main}
\end{document}